\documentclass[preprint,11pt,nobibnotes,showpacs,showkeys,a4paper,aps,prb]%
{revtex4}
\usepackage[dvips]{graphicx,graphics,color,epsfig}
\usepackage{amsmath,amsfonts,amssymb}
\usepackage{verbatim}
\usepackage{latexsym}
\usepackage{dcolumn}
\usepackage{bm}

\begin{document}
\title{Sterile neutrinos influence on oscillation characteristics of active
neutrinos at short distances in the generalized model of neutrino mixing}
\author{V. V. Khruschov$^{\ast,\dagger,\$}$ and
S. V. Fomichev$^{\ast,\ddagger,\P}$}
\affiliation{$^{\ast}$National Research Centre Kurchatov Institute,
Academician Kurchatov Place 1, Moscow, 123182 Russia}
\affiliation{$^{\dagger}$Center for Gravitation and Fundamental Metrology,
VNIIMS, Ozernaya Str. 46, Moscow, 119361 Russia}
\affiliation{$^{\ddagger}$Moscow Institute of Physics and Technology (State
University), Institutskii Lane 9, Dolgoprudnyi, Moscow Region, 141700 Russia}
\affiliation{$^{\$}$khruschov\_vv@nrcki.ru}
\affiliation{$^{\P}$fomichev\_sv@nrcki.ru}

\begin{abstract}
A phenomenological model with active and sterile neutrinos is used for
calculations of neutrino oscillation characteristics at the normal mass
hierarchy of active neutrinos. Taking into account the contributions of
sterile neutrinos, appearance and survival probabilities for active neutrinos
are calculated. Modified graphical dependencies for the probability of
appearance of electron neutrinos/antineutrinos in muon neutrino/antineutrino
beams as a function of the ratio of the distance to the neutrino energy and
other model parameters are obtained. It is shown that in the case of a certain
type mixing between active and sterile neutrinos it is possible to clarify
some features of the anomalies of neutrino data at short distances. A new
parametrization for a particular type mixing matrix of active and sterile
neutrinos that takes into account the additional sources of CP violation is
used. The comparison with the existing experimental data is performed and,
with using this knowledge, the estimates of some model parameters are found.
The theoretical results obtained for mixing of active and sterile neutrinos
can be applied for interpretation and prediction of results of ground-based
experiments on search of sterile neutrinos as well as for the analysis of some
astrophysical data.
\end{abstract}


\maketitle

\section{Introduction}
\label{Section_Introduction}

It is well known that oscillations of solar, atmospheric, reactor and
accelerator active neutrinos (AN) can be attributed to mixing of three mass
states of neutrinos. The mixing of these states \cite{Bilenky1977} puts into
operation with the Pontecorvo--Maki--Nakagawa--Sakata matrix
$U_{\rm PMNS}\equiv U\equiv V\!P$, so that $\psi_a^L=\sum_iU_{ai}\psi_i^L$,
where $\psi_{a,i}^L$ are left chiral fields with flavor $a$ or mass $m_i$,
$a=\{e,\mu,\tau\}$ and $i=\{1,2,3\}$. For three active neutrinos, the matrix
$V$ is expressed in the standard parametrization \cite{PDG} via the mixing
angles $\theta_{ij}$ and the CP-phase, namely, the phase
$\delta\equiv\delta_{\rm CP}$ associated with CP violation in the lepton
sector for Dirac or Majorana neutrinos, and
$P={\rm diag}\{1,e^{i\alpha},e^{i\beta}\}$, where
$\alpha\equiv\alpha_{\rm CP}$ and $\beta\equiv\beta_{\rm CP}$ are phases
associated with CP violation only for Majorana neutrinos. In the atmospheric,
solar, reactor and accelerator oscillation experiments it is impossible to
measure $\alpha_{\rm CP}$ and $\beta_{\rm CP}$ and attribute neutrinos to
Majorana or Dirac type of particles. Meanwhile, obtained experimental results
made it possible to establish a breaking of conservation for lepton numbers
$L_e$, $L_{\mu}$ and $L_{\tau}$.

With the help of high-precision experimental data, the values of the mixing
angles and the differences of the neutrino masses in square $\Delta m_{21}^2$
and $\Delta m_{31}^2$ were found \cite{PDG,Esteban2017} (where
$\Delta m_{ij}^2=m_i^2-m_j^2$). For these neutrino masses in square
differences only absolute value of $\Delta m_{31}^2$ is known, therefore, the
absolute values of the neutrino masses can be ordered by two ways, namely, as
$m_1<m_2<m_3$ or $m_3<m_1<m_2$. These cases are called as normal hierarchy
(NH) and as inverse hierarchy (IH) of the neutrino mass spectrum,
respectively. Including nonzero neutrino masses leads to the Minimally
Extended Standard Model (MESM) instead of the Standard Model (SM). Although
the value of CP-phase $\delta_{\rm CP}$ is not yet definitively determined
experimentally, in a number of papers its estimate was obtained (for example,
see Refs.~\onlinecite{Esteban2017,KhruFom2016,Petkov,Kabe}). For the NH-case
of the mass spectrum of active neutrinos we have $\sin\delta_{\rm CP}<0$ and
$\delta_{\rm CP}\approx -\pi/2$. If we take into account the restrictions on
the sum of the neutrino masses from cosmological observations \cite{Wang} and
the results of the T2K experiment \cite{Kabe}, then the NH-case of the
neutrino mass spectrum turns out to be preferable. So, in carrying out further
numerical calculations we restrict ourselves to the NH-case only, assuming
$\delta_{\rm CP}\equiv\delta_1=-\pi/2$.

At the same time, there are indications to anomalies of neutrino fluxes for
some processes that can not be explained with using oscillation parameters
only for three active neutrinos. These anomalies include LSND (or accelerator)
anomaly \cite{Atha1996,Agu2001,Agu2013,Agu2018}, reactor
\cite{Mu2011,Me2011,Hu2011,Ko,ale18,ser18} and gallium (calibration)
\cite{Abdu2009,Kae2010,Giunti2013} anomalies. The anomalies manifest
themselves at short distances (more precisely, at distances $L$ such that the
numerical value of the parameter $\Delta m^2 L/E$, where $E$ is the neutrino
energy, is of the order of unity). In the LSND anomaly, an excess of the
electron antineutrinos in beams of muon antineutrinos in comparison with the
expected value according to the MESM is observed. Similar results were
observed in the MiniBooNE experiments for electron neutrinos and antineutrinos
\cite{Agu2013,Agu2018}. Deficit of reactor electron antineutrinos at short
distances is called as reactor anomaly, while the deficit of electron
neutrinos from a radioactive source occurred at calibration of detectors for
the SAGE and GALLEX experiments is commonly called as gallium anomaly. In
other words, data on anomalies refer to both the appearance of electron
neutrinos or antineutrinos in beams of muon neutrinos or antineutrinos,
respectively, and to the disappearance of electron neutrinos or antineutrinos.
These three types of the shot-baseline (SBL) neutrino anomalies, for which
there are indications at present, may be explained by the existence of one or
two new neutrinos that do not interact directly with the gauge bosons of the
SM, that is sterile neutrinos (SN). The characteristic mass scale of sterile
neutrino used for explanation of the SBL anomalies is about $1$~eV.

Today, in addition to SN, other new neutral particles are present in many
models that are beyond the framework of SM such as supersymmetric models,
grand unification theories, different phenomenological models, etc. The using
of some new particles is associated with necessity to explain a number of
phenomena in cosmology, astrophysics and particle physics that are difficult
to explain in terms of SM particles only. As an example we point to neutral
fermions with large masses (heavy neutrinos (HN)) and dark matter particles
(DMP), of which, perhaps, dark matter (DM) consists, at that SN and HN can
be components of DMP. Almost all characteristics of DMP remain unknown now
\cite{Vogelsberger2014,Demi2014}. DM is not of the baryon nature and, in its
turn, probably consists of cold dark matter (CDM), warm dark matter (WDM) and
hot dark matter (HDM). Models with a multicomponent dark matter are often used
to describe various structures with different scales in the Universe. More
details about properties of DMP, including SN and HN, can be found in numerous
papers (see, for example,
Refs.~\onlinecite{Demi2014,Abazajian2012,Abe2017,Kusenko2009,Liao}).

The mass values of SN, HN and DMP belong to a wide range from $10^{-6}$~eV to
$10^{16}$~GeV \cite{deGouvea2005,Gorbunov2014,Drewes2015}. Their interactions
with MESM particles can be realized by means of new scalar, pseudoscalar or
vector bosons, in some cases new vector bosons can be mixed with the photon
and/or the $Z$-boson. The latest data obtained on studying the Cosmic
Microwave Background with the help of a number of cosmological models lead to
a restriction on the number of new relativistic particles that were in the
thermodynamic equilibrium in the era of plasma recombination in the early
Universe \cite{Ade2014,Abaz2017,Gerbino}. So, it is convenient to associate a
mass scale between light and heavy neutrinos with the values of temperature,
which are typical for the recombination epoch and belong to the eV-range. The
maximum value of this temperature is $R_{\infty}hc \approx 13.6$~eV
($R_{\infty}$ is the Rydberg constant). That is, neutrinos with masses smaller
than or equal to $13.6$~eV may be called as light neutrinos (LN), while
neutrinos with masses large than $13.6$~eV may be called as heavy neutrinos
(HN). Then the LN will include well-known active neutrinos $\nu_e$,
$\nu_{\mu}$ and $\nu_{\tau}$, and also perhaps new light SN. If one takes into
account the possibility of realization of non-standard models, where sterile
LN did not thermalize in the plasma of the early Universe, then LN effect in
cosmological data should be suppressed (see, for example,
Refs.~\onlinecite{Hannestad2014,Chu2015}). HN with masses less than half of
the mass of the $Z$-boson are SN, while more heavy neutrinos, in principle,
can interact directly with $Z$-boson, as, for example, a hypothetical heavy
neutrino from the fourth generation or WIMPs.

At present, intensive searches are carried out for light SNs with masses of
the order of $1$~eV, which are proposed for explanation of the SBL anomalies.
It is expected that in the coming several years it will be possible to confirm
or deny the existence of such anomalies and light SN (see, for example,
Refs.~\onlinecite{Abazajian2012,Gorbunov2014,Gav2017,Gav2018,Bel2013,Ser2015}).
Besides, HN with masses from several keV to several TeV are often used to
explain some astrophysical data \cite{Demi2014,Mar2016,Arg2017}. As the
existence of SN and HN goes beyond MESM, there have been proposed
phenomenological models for prediction of their characteristics, as well as
effects due to them (see, for example, Refs.~\onlinecite{%
KhruFom2016,Abe2017,Canetti2013,Conrad2013,Zysina2014,KhruFom2017}). For
instance, in Ref.~\onlinecite{PAZH2015} the effect of increasing of sterile
neutrinos yield in a high-density medium when the ratio of the number of
neutrons to the number of protons approaches to two was considered in detail.
This effect can have impact on the characteristics of fluxes of active
neutrinos in supernovae \cite{Warren,PAZH2016}. Phenomenological models with
additional SN and HN are usually denoted as (3+$N$) models, or, in detail,
as ($k$+3+$n$+$m$) models, where $k$ is the number of new neutrinos with
masses less than masses of active neutrinos, and $n$ and $m$ are the numbers
of new neutrinos with masses higher and considerably higher, respectively,
than masses of the active neutrinos \cite{Bilenky1977,Abazajian2012,%
Zysina2014,KhruFom2017,Schwetz2011,Kopp2013,Gariazzo2017,Bilenky}.

In this paper, we present a phenomenological (3+3) model \cite{KhF-mephi} with
three active neutrinos and three SN and consider the effects of SN, which
appear in oscillation characteristics of active neutrinos with energies of the
order of MeV or dozens of MeV at small distances (of the order of several
meters or tens of meters, corresponding to the SBL experiments). In
Section~\ref{Section_OscillationModel}, the main concepts of our (3+3) model
(to be exact, the (3+1+2) model) based on the results obtained earlier
\cite{KhruFom2016,KhruFom2017} are given in detail, while in
Section~\ref{Section_CalculationResults} the results of detailed calculations
of the oscillation characteristics of active neutrinos at small distances with
account of effects of SN are presented. Calculations were carried out with the
use of new parametrization for the generalized mixing matrix of LN and HN at
selected test values of the model parameters. By introducing additional
CP-phases in the framework of our model, one can explain asymmetry of the
yield of electron neutrinos and antineutrinos in beams of muon neutrinos and
antineutrinos, respectively. The comparison is made with the simplest (3+1)
model, where in the main approximation the asymmetry is absent. The results of
the performed SBL experiments are taken into account and on this basis
tentative estimates of some model parameters are made. In the final
Section~\ref{Section_Conclusion} it is noted that the obtained results can
help to explain the available experimental data on the LSND and MiniBooNE
anomaly and reactor and gallium anomalies, as well as to interpret both
expected data of SBL experiments on the search of sterile neutrinos and some
astrophysical data.

\section{Basic concepts of the phenomenological (3+1+2) model of light and
heavy neutrinos}
\label{Section_OscillationModel}

The (3+$N$) or (3+$m$+$n$) phenomenological neutrino models can be used to
describe the SBL anomalies, as well as some astrophysical data, where $N=m+n$
is the number of additional neutrinos, which, in principle, can be arbitrary
(see, for example,
Refs.~\onlinecite{Bilenky1977,Abazajian2012,Bilenky,KhF-mephi}). It is
desirable that the number of new neutrinos would be minimal, so the most
common are the (3+1) and (3+2) models. However, if we apply the principle of
extended symmetry of weak interactions, then, for example, for the left-right
symmetry it is necessary in this case to consider (3+3) models
\cite{KhruFom2016,Conrad2013,Zysina2014,KhruFom2017}. So, below we consider a
(3+1+2) model that can be used to describe effects of light and heavy SN. This
model includes three active neutrinos $\nu_a$ ($a=e,\mu,\tau$) and three new
neutrinos: a sterile neutrino $\nu_s$, a hidden neutrino $\nu_h$ and a dark
neutrino $\nu_d$. We use a characteristic order of mass values of $\nu_s$,
$\nu_h$ and $\nu_d$ as well as a new parametrization of the mixing matrix,
so our  model can be considered as a generalization of the (3+3) model, which
was studied in Refs.~\onlinecite{KhruFom2016,Zysina2014,KhruFom2017}.

In order to take into account the contributions of light and heavy SN to the
oscillation characteristics of active neutrinos, it will be considered the
$6\!\times\!6$ mixing matrix, which can be called as the generalized mixing
matrix $U_{\rm mix}$, or the generalized Pontecorvo--Maki--Nakagawa--Sakata
matrix $U_{\rm GPMNS}\equiv U_{\rm mix}$ \cite{KhruFom2016,KhruFom2017}. This
matrix can be represented as the matrix product $V\!P$, where $P$ is a
diagonal matrix with Majorana CP-phases $\phi_i$, $i=1,\dots,5$, namely,
$P={\rm diag}\{1,e^{i\phi_1},\dots,e^{i\phi_5}\}$. Below we will consider only
the particular type of matrix $V$. Keeping continuity of the notations, we
will denote fifteen Dirac CP-phases as $\delta_i$ and $\kappa_j$, and twenty
one mixing angles as $\theta_i$ and $\eta_j$, where
$\delta_1\equiv\delta_{\rm CP}$, $\theta_1\equiv\theta_{12}$,
$\theta_2\equiv\theta_{23}$ and $\theta_3\equiv\theta_{13}$.

For the compactness of the formulas, we introduce the symbols $h_s$ and
$h_{i'}$ for generalized left flavor fields and generalized left mass fields,
respectively. As $s$ we will use a set of indices that allocate $\nu_s$,
$\nu_h$ and $\nu_d$ fields among $h_s$, and as $i'$ we will use a set of
indices $4$, $5$ and $6$. The common $6\!\times\!6$ mixing matrix
$U_{\rm mix}$ can then be expressed through $3\!\times\!3$ matrices $R$, $T$,
$V$ and $W$ as follows
\begin{equation}
\left(\begin{array}{c}\nu_a\\ h_s \end{array}\right)=
U_{\rm mix}\left(\begin{array}{c}\nu_i\\ h_{i'}\end{array}\right)\equiv
\left(\begin{array}{cc}R&T\\ V&W\end{array}\right)
\left(\begin{array}{c}\nu_i\\ h_{i'}\end{array}\right).
\label{eq_Umix}
\end{equation}
We represent the matrix $R$ in the form of
$R=U_{\rm PMNS}+\Delta U_{\rm PMNS}$, where the matrix $\Delta U_{\rm PMNS}$,
as well as the matrix $T$ in equation~(\ref{eq_Umix}) should be small as
compared with the matrix $U_{\rm PMNS}$. For the convenience of quantitative
estimates of arising corrections to mixing between active neutrinos due to
SN, we will put $\Delta U_{\rm PMNS}=-\epsilon U_{\rm PMNS}$, where $\epsilon$
is a small value, which can be presented as $\epsilon=1-\varkappa$. Then the
matrix $R$ will be represented as $R=\varkappa U_{\rm PMNS}$, that is, it will
be proportional to the known unitary $3\!\times\!3$ mixing matrix of active
neutrinos ($U_{\rm PMNS}U_{\rm PMNS}^+=I$). Then we will use the notation
$U_{\rm PMNS}\equiv U$.

Thus, when choosing the appropriate normalization, the active neutrinos mix,
as it should be in the MESM, according to Pontecorvo--Maki--Nakagawa--Sakata
matrix. Bearing in mind that, in accordance with data available due to
astrophysical and laboratory measurements, the mixing between active and new
neutrinos is small, we choose the matrix $T$ as $T=\sqrt{1-\varkappa^2}\,a$,
where $a$ is an arbitrary unitary $3\!\times\!3$ matrix, that is, $aa^+=I$.
The matrix $U_{\rm mix}$ can now be written in the form of
\begin{equation}
U_{\rm mix}=\left(\begin{array}{cc}R&T\\ V&W\end{array}\right)\equiv
\left(\begin{array}{cc}\varkappa U&\sqrt{1-\varkappa^2}\,a\\
\sqrt{1-\varkappa^2}\,bU&\varkappa c \end{array}\right),
\label{eq_Utilde}
\end{equation}
where $b$ is also an arbitrary unitary $3\!\times\!3$ matrix ($bb^+=I$), and
$c=-ba$. With these conditions, the matrix $U_{\rm mix}$ will be unitary
($U_{\rm mix}U_{\rm mix}^+=I$). For the matrix $U_{\rm mix}$ we will consider,
taking into account additional physical reasons, only some particular cases,
but not the most common form. In particular, we will use the following
matrices $a$ and $b$:
\begin{equation}
a=\left(\begin{array}{lcr}\,\,\,\,\,\cos\eta_2 & \sin\eta_2 & 0\\
-\sin\eta_2 & \cos\eta_2 & 0\\
\qquad 0 & 0 & e^{-i\kappa_2}\end{array}\right),\quad
b=-\left(\begin{array}{lcr}\,\,\,\,\,\cos\eta_1 & \sin\eta_1 & 0\\
-\sin\eta_1 & \cos\eta_1 & 0\\
\qquad 0 & 0 & e^{-i\kappa_1}\end{array}\right),
\label{eq_matricesab}
\end{equation}
where $\kappa_1$ and $\kappa_2$ are mixing phases for active and sterile
neutrinos, whereas $\eta_1$ and $\eta_2$ are their mixing angles. The matrix
$a$ in the form of equation~(\ref{eq_matricesab}) was proposed in
Ref.~\onlinecite{KhruFom2016}. In order to make our calculations more
specific, we will use the following test values for new mixing parameters:
\begin{equation}
\kappa_1=\kappa_2=-\pi/2,\quad \eta_1=5^{\circ},\quad \eta_2=\pm 30^{\circ},
\label{eq_etakappa}
\end{equation}
and assume that the small parameter $\epsilon$ satisfies at least the
condition $\epsilon\lesssim 0.03$.

Note that the mixing matrix in the form of equation~(\ref{eq_Utilde}) is more
general in comparison with the mixing matrix $\widetilde{U}$ that was proposed
and used in Ref.~\onlinecite{KhruFom2016}. Indeed, in that paper the
$3\!\times\!3$ matrix $c$ was reduced to a diagonal matrix, in fact to a phase
factor $e^{i\phi}$, and also there was no additional independent mixing angle
$\eta_1$, since the matrix $b$ was chosen proportional to matrix $a^+$. In the
version of neutrino mixing discussed here, that corresponds to
equations~(\ref{eq_Utilde}) and (\ref{eq_matricesab}), there are more
possibilities to describe the various contributions of SN.

The neutrino masses will be given by a normally ordered set of values
$\{m\}=\{m_i,m_{i'}\}$. For active neutrinos we will use the neutrino mass
estimations, which were proposed in
Refs.~\onlinecite{KhruFom2016,Zysina2014,PAZH2016} for NH-case (in units of
eV) and which do not contradict to the known experimental data:
\begin{equation}
m_1\approx 0.0016, \quad m_2\approx 0.0088, \quad m_3\approx 0.0497\,.
\label{eq_activmasses}
\end{equation}
The values of the mixing angles $\theta_{ij}$ of active neutrinos that
determine the Pontecorvo--Maki--Nakagawa--Sakata mixing matrix will be taken
from relations $\sin^2\theta_{12}\approx 0.297$,
$\sin^2\theta_{23}\approx 0.425$ and $\sin^2\theta_{13}\approx 0.0215$,
which are obtained from the processing of experimental data and given in
Ref.~\onlinecite{PDG}.

In what follows we consider two variants of the (3+1+2) model, which differ by
the sterile neutrino mass $m_4$. The values of this mass are selected on the
basis of experimental results of MiniBooNE, NEOS, DANSS and Neutrino-4
experiments \cite{Agu2018,Ko,ale18,ser18}, which point to the values of $0.2$,
$1.14$, $1.18$ and $2.65$~eV, respectively, as at the best fit for sterile
neutrino mass. Taking into account the considerable dispersion of these
results, we choose two $m_4$ values, $0.55$ and $1.1$~eV, as possible test
values in the framework of our model.

On the other hand, if to select, together with the mass value of the light
neutrino $m_4 \sim 1$~eV, the mass value $m_6$ associated with particle
$\nu_d$ as about $10$~keV, it becomes possible to explain the appearance of
anomalies at short distances in neutrino data \cite{Gariazzo2017}, as well as
the registration of the line $3.55$~keV in the gamma spectra of some
astrophysical sources \cite{Bulb2014,Boya2014,Horiuchi}. Note that sterile
neutrinos with masses of the order of $1$~keV are also used for interpretation
of some astrophysical data, so we choose the mass value $m_5$ associated with
particles $\nu_h$ as about $1$~keV. Thus, let us to consider the mass option,
which can be designated as ``Light Mass Option" (LMO), in two variants:
\begin{equation}
\{m\}_{\rm LMO1}=\{1.1,\,1.5\!\times\!10^3,\,7.5\!\times\!10^3 \}.
\label{eq_LMO1}
\end{equation}
\begin{equation}
\{m\}_{\rm LMO2}=\{0.55,\,1.5\!\times\!10^3,\,7.5\!\times\!10^3 \}.
\label{eq_LMO2}
\end{equation}

The probability amplitudes for propagation of neutrino flavors can be found by
solution of well-known equations (see, for example,
Ref.~\onlinecite{ble,KhruFom2016}). Moreover, with the help of these
equations, analytical expressions for transition probabilities between
different flavors in neutrino/antineutrino beams in vacuum as a function of
the distance from the neutrino source can be obtained \cite{Bilenky}, which
are also used in the current paper in calculations to control the results
obtained by numerical solution of the equations.

For three active neutrinos, almost always ultrarelativistic, these equations
have the form
\begin{equation}
i\partial_{r}\left(\begin{array}{c} a_{e}\\ a_{\mu}\\ a_{\tau}\end{array}
\right)=H\left(\begin{array}{c} a_{e}\\ a_{\mu}\\
a_{\tau}\end{array}\right),
\label{eq_active_n_amplitudes}
\end{equation}
where the matrix $H$ is expressed using the matrix $U_{\rm PMNS}\equiv U$ in
the form of
\begin{equation}
H=\frac{U}{2E}\!\left(\begin{array}{ccc}m_{1}^{2}-m_{0}^{2} & 0 & 0 \\
0 & m_{2}^{2}-m_{0}^{2} & 0 \\ 0 & 0 & m_{3}^{2}-m_{0}^{2}
\end{array}\right)\!U^{+}.
\label{eq_activeH}
\end{equation}
Here $m_0$ is the smallest value among three neutrino masses $m_1$, $m_2$ and
$m_3$, and $E$ is the neutrino energy. In what follows, as a basic case it
will be used here the simplest conventional approach for neutrino oscillations
that is based on the plane-wave neutrino states. The more consistent approach
with wave packets of the neutrino states (see, e.g., Ref.~\onlinecite{akh}),
which takes into account some coherence limitations, can also be considered in
further elsewhere.

In the plane-wave approximation, the neutrinos possess equal momentums that
leads to the diagonal neutrino energy matrix $\Delta_{E}$ in the form of
\begin{equation}
\Delta_{E}={\rm diag}\{E_1-E_0,\,E_2-E_0,\,\ldots,\,E_{6}-E_0\}\,,
\label{eq_DeltaE}
\end{equation}
where $E_i=\sqrt{p^2+m_i^2}$, $m_i$ ($i=1,2,\ldots 6$) are the neutrino masses
and $m_0$ is the smallest mass among $m_i$. The momentum $p$ can be
related to the energy $E\approx p$ of ultrarelativistic active neutrinos. In
the ultrarelativistic limit for all neutrinos, in place of the matrix
$\Delta_{E}$ it is possible to use the matrix $\Delta_{m^2}$ of the
differences of the squares of neutrino masses, which in the general case of
3+$N$ flavors is defined as
\begin{equation}
\Delta_{m^2}={\rm diag}\{m_1^2-m_0^2,\,m_2^2-m_0^2,\,\ldots,
\,m_{3+N}^2-m_0^2\}\,.
\label{eq_minsquarematrix}
\end{equation}
Then it is necessary to solve the following equations for neutrino
propagation, similar to the equations~(\ref{eq_active_n_amplitudes}) and
(\ref{eq_activeH}) for active neutrinos:
\begin{equation}
i\partial_{r}\left(\begin{array}{c} a_{a}\\ a_{s} \end{array}\right)=
\frac{U_{\rm mix}}{2E}\Delta_{m^2}U^{+}_{\rm mix}
\left(\begin{array}{c} a_{a}\\ a_{s} \end{array}\right),
\label{eq_neutrino_m2}
\end{equation}
where $U_{\rm mix}$ is the unitary $6\!\times\!6$ neutrino mixing matrix given
by equations (\ref{eq_Utilde})--(\ref{eq_matricesab}) and $r=r_a\approx ct$ is
the distance traveled by active neutrinos (it was assumed above that $c=1$).
For antineutrinos, the equations have the form
\begin{equation}
i\partial_{r}\left(\begin{array}{c} {a}_{\overline{a}}\\ {a}_{\overline{s}}
\end{array}\right)=
\frac{U_{\rm mix}^*}{2E}\Delta_{m^2}U^T_{\rm mix}\left(\begin{array}{c}
{a}_{\overline{a}}\\ {a}_{\overline{s}} \end{array}
\right),
\label{eq_antineutrino_m2}
\end{equation}
where $\ast$ means complex conjugation. Solving these equations for certain
values of the parameters, one can find the survival probabilities, and also
probabilities of appearance and disappearance of neutrinos or antineutrinos
of any flavor as functions of the neutrino/antineutrino energy and the
distance from the source.

From equations (\ref{eq_minsquarematrix})--(\ref{eq_antineutrino_m2}),
analytical expressions for the probabilities of transitions between different
flavors of neutrinos/antineutrinos in vacuum as a function of the distance $L$
from the source can be obtained. If $\widetilde{U}\equiv U_{\rm mix}$ is a
generalized $6\!\times\!6$ mixing matrix in the form given by equation
(\ref{eq_Utilde}), and if we use the notation
$\Delta_{ki}\equiv\Delta m_{ik}^2L/(4E)$, then, following by
Ref.~\onlinecite{Bilenky}, it is possible to calculate the transition
probabilities from $\nu_{\alpha}$ to $\nu_{\alpha^{\prime}}$ or from
$\overline{\nu}_{\alpha}$ to $\overline{\nu}_{\alpha^{\prime}}$ by the formula
\begin{eqnarray}
P(\nu_{\alpha}(\overline{\nu}_{\alpha})\rightarrow\nu_{\alpha^{\prime}}
(\overline{\nu}_{\alpha^{\prime}}))=\delta_{\alpha^{\prime}\alpha}
&-4\sum_{i>k}{\rm Re}(\widetilde{U}_{\alpha^{\prime} i}
\widetilde{U}_{\alpha i}^{\ast}\widetilde{U}_{\alpha^{\prime} k}^{\ast}
\widetilde{U}_{\alpha k})\sin^2\Delta_{ki}\,\nonumber \\
&\pm2\sum_{i>k}{\rm Im}(\widetilde{U}_{\alpha^{\prime} i}
\widetilde{U}_{\alpha i}^{\ast}\widetilde{U}_{\alpha^{\prime} k}^{\ast}
\widetilde{U}_{\alpha k})\sin 2\Delta_{ki}\,,
\label{eq_Bilenky}
\end{eqnarray}
where the upper sign $(+)$ corresponds to neutrino transitions
$\nu_{\alpha}\rightarrow\nu_{\alpha^{\prime}}$, while the lower sign $(-)$
corresponds to antineutrino transitions
$\overline{\nu}_{\alpha}\rightarrow\overline{\nu}_{\alpha^{\prime}}$. Note,
that the flavor indices $\alpha$ and $\alpha^{\prime}$ (also as summation
indices $i$ and $k$ over massive states) apply to all neutrinos, that is to
active, sterile and heavy neutrinos. Moreover, as follows from equation
(\ref{eq_Bilenky}), the relation $P(\nu_{\alpha}\rightarrow\nu_{\alpha})\equiv
P(\overline{\nu}_{\alpha}\rightarrow\overline{\nu}_{\alpha})$ is fulfilled
exactly as a consequence of the CPT-invariance \cite{Bilenky}.

To check the accuracy of the numerical results obtained on the basis of
equations (\ref{eq_neutrino_m2}) and (\ref{eq_antineutrino_m2}) taking into
account the subtle effects of the possible existence of SN and HN,
calculations were also performed with the help of precise analytical
expressions (\ref{eq_Bilenky}). Moreover, the probabilities of the processes
of interest to us, namely, the processes of appearance of electron
neutrinos/antineutrinos in a beam of muon neutrino/antineutrino depend only
on the first two rows of the matrix $U_{\rm mix}$ and can be explicitly
written (see the Appendix). From expressions~(\ref{eqA1})--(\ref{eqA8}) it is
clearly seen that some parameters of the model that were introduced before
for the sake of generality, namely, the mixing phases $\kappa_1$ and
$\kappa_2$, the mixing angle $\eta_1$ and the heavy neutrino mass $m_6$ are
not involved in the processes of appearance of electron
neutrinos/antineutrinos in a beam of muon neutrinos/antineutrinos. Only
parameter $\epsilon$, masses $m_4$ and $m_5$ of two sterile/heavy neutrinos
and the mixing angle $\eta_2$, along with the mixing parameters of three
active neutrinos are responsible for these processes. This is also true for
processes of survival of electron neutrinos/antineutrinos that are responsible
for the gallium and reactor anomalies.

\section{Numerical results for oscillation characteristics of active neutrinos
with the account of their mixing with SN}
\label{Section_CalculationResults}
\subsection{Computation of oscillation characteristics at the test values of
model parameters}
\label{Subection_TestValuesOfModelParameters}

In this paper, we mainly focus on the possibility of describing, in the
framework of the considered model, the anomalies found in the data of LSND and
MiniBooNE experiments on oscillations of accelerator muon neutrinos and
antineutrinos, which were subsequently tested and will still be tested in
accelerator experiments \cite{Gariazzo,Kudenko,Agu2018}. It refers to data on
the disappearance of muon neutrinos and antineutrinos and the appearance of
electron neutrinos and antineutrinos in the processes ${\nu}_{\mu}\to{\nu}_e$
and $\overline{\nu}_{\mu}\to\overline{\nu}_e$. The ratio of the distance $L$
traveled by the neutrino before detection to the neutrino energy $E$ is
typically a few meters per one ${\rm MeV}$. Attempts of the simultaneous
description of all the data in these processes leads to difficulties. In
particular, the problem associated with different values of the excess of the
output of ${\nu}_e$ and $\overline{\nu}_e$ in the MiniBooNE experiment may be
resolved under the condition of CP violation \cite{Mal2007,Pal2005,Kar2007}.

Note that the reactor and gallium anomalies manifesting themselves in neutrino
data as the disappearance of electron neutrinos and antineutrinos can be
described in the framework of the model considered in this paper with the new
parametrization of the mixing matrix by selecting the value of the parameter
$\epsilon$. To describe these anomalies, it is sufficient to choose the
appropriate value less than unity and corresponding to the experimental data
for the parameter $\varkappa=1-\epsilon$, which is present in our
parametrization of the mixing matrix (see equation~(\ref{eq_Utilde})) that
naturally leads to the deficit of electron neutrinos and antineutrinos. The
status of the reactor anomaly with allowance for the recently discovered
excess of the number of antineutrinos in comparison with the model
calculations in the $5$~MeV range and confirmation of the possible existence
of a light SN with a mass of about $1$~eV see, for example, in
Refs.~\onlinecite{Dentler,Dan,Gari}.

\begin{figure}
\includegraphics[width=0.99\textwidth]{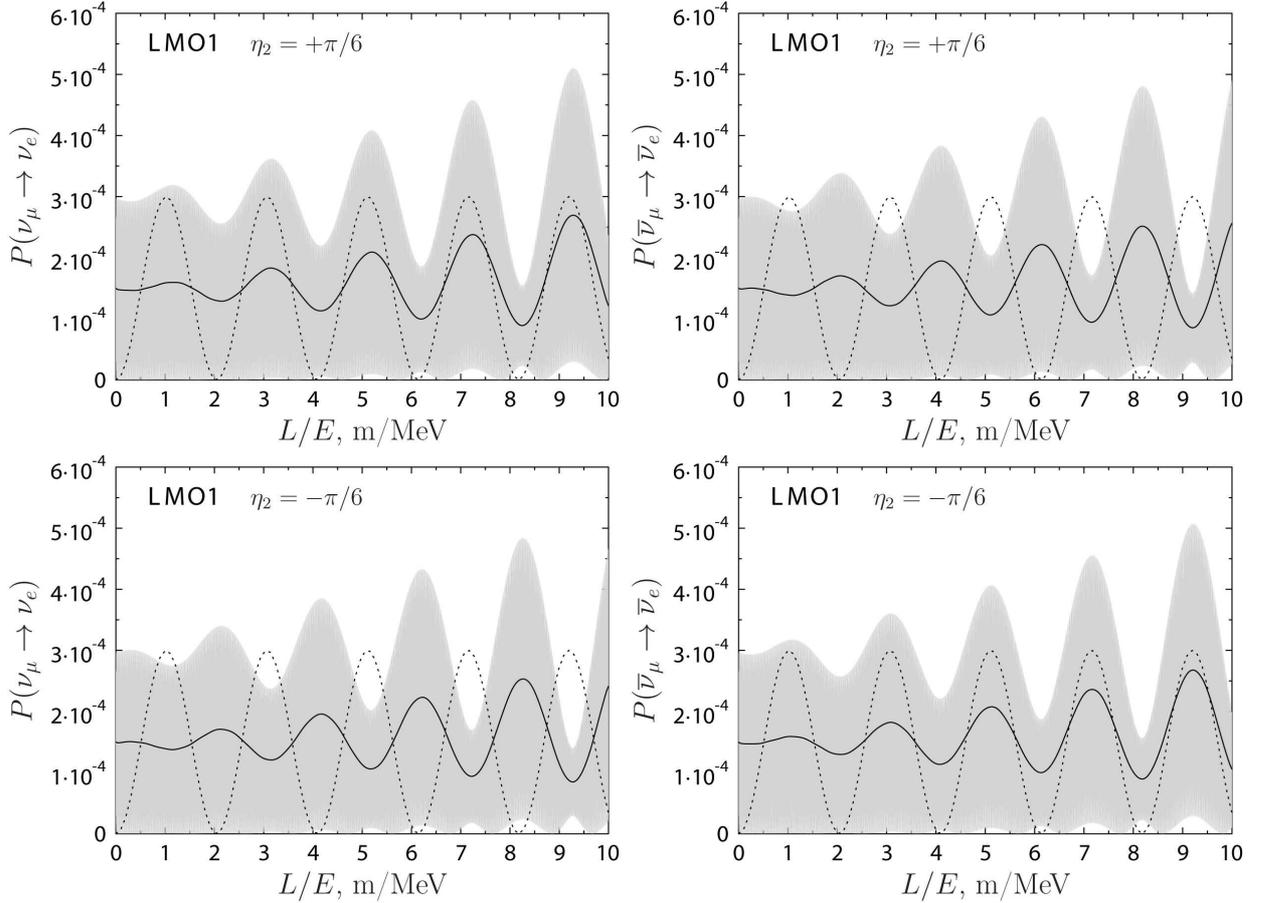}
\caption{The probability of appearance of electron neutrinos (left panels) and
antineutrinos (right panels) versus the ratio of the distance $L$ from the
source to the neutrino energy $E$ in the beams of muon neutrinos and
antineutrinos, respectively. The value $\epsilon=0.01$ of the coupling
constant of active and sterile neutrinos is taken for the case of the mixing
matrix considered in this paper
(equations~(\ref{eq_Utilde})--(\ref{eq_matricesab})), for $\eta_2=\pm\pi/6$,
and for the LMO1 version (equation~(\ref{eq_LMO1})) of the mass values of
sterile neutrinos. The gray region arises as a result of exact calculations
due to fast oscillations caused by the presence in the model of fifth sterile
neutrino with mass of the order of $1$~{\rm keV}, while the solid curves show
probability values averaged over small-scale spatial oscillations. The dotted
curves show probability values calculated in the simplest approximation of the
(3+1) model and the two-neutrino mixing with $\sin^2(2\theta)=0.0003$ and
$\Delta m^2_{41}=1.21~{\rm eV}^2$.}
\label{Fig_LMO1}
\end{figure}
\begin{figure}
\includegraphics[width=0.99\textwidth]{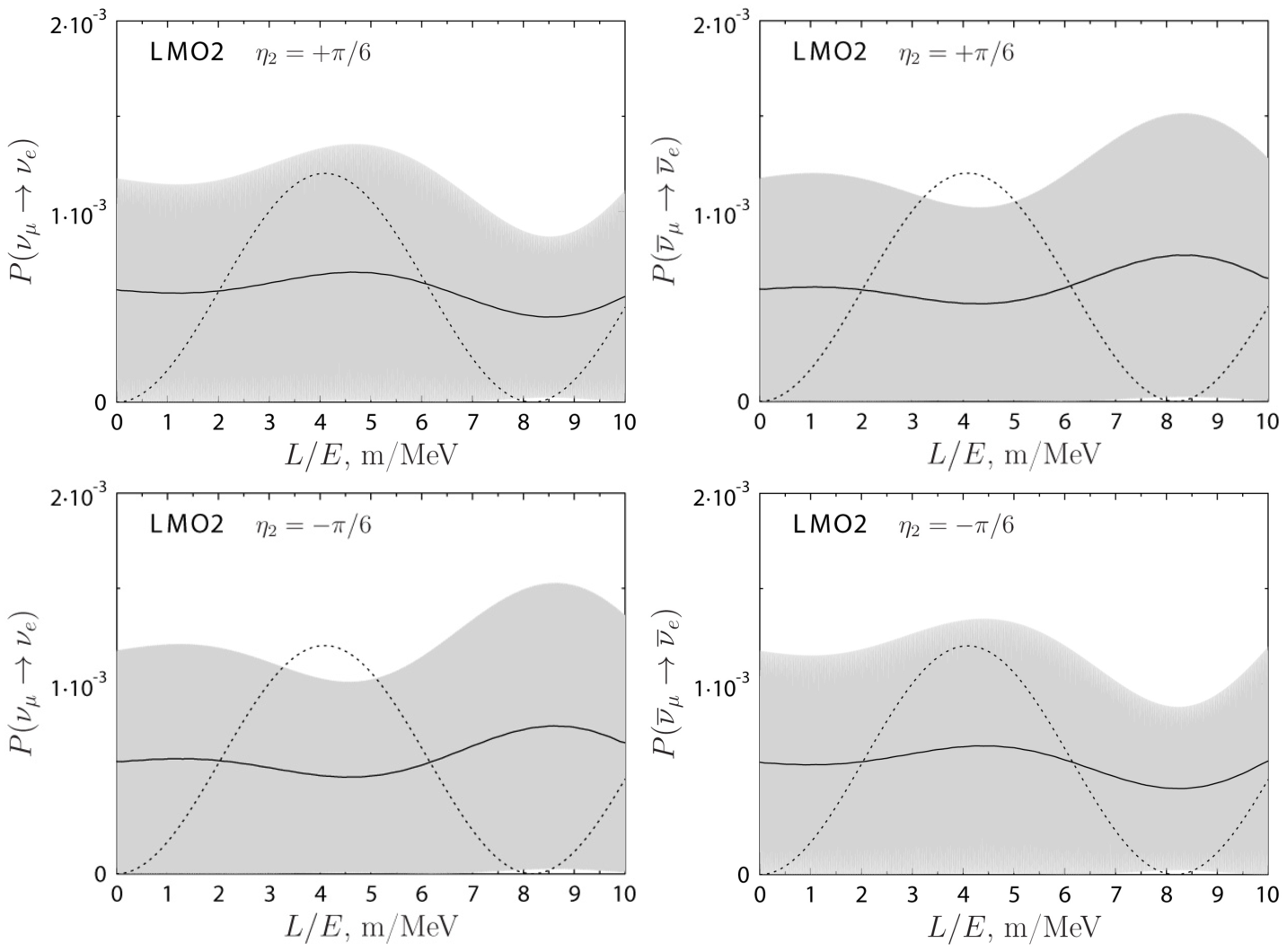}
\caption{The probability of appearance of electron neutrinos (left panels) and
antineutrinos (right panels) versus the ratio of the distance $L$ from the
source to the neutrino energy $E$ in the beams of muon neutrinos and
antineutrinos, respectively. The value $\epsilon=0.02$ of the coupling
constant of active and sterile neutrinos is taken for the case of the mixing
matrix considered in this paper
(equations~(\ref{eq_Utilde})--(\ref{eq_matricesab})), for $\eta_2=\pm\pi/6$,
and for the LMO2 version (equation~(\ref{eq_LMO2})) of the mass values of
sterile neutrinos. The gray region arises as a result of exact calculations
due to fast oscillations caused by the presence in the model of fifth sterile
neutrino with mass of the order of $1$~{\rm keV}, while the solid curves show
probability values averaged over small-scale spatial oscillations. The dotted
curves show probability values calculated in the simplest approximation of the
(3+1) model and the two-neutrino mixing with $\sin^2(2\theta)=0.0012$ and
$\Delta m^2_{41}=0.3~{\rm eV}^2$.}
\label{Fig_LMO2}
\end{figure}
\begin{figure}
\includegraphics[width=0.99\textwidth]{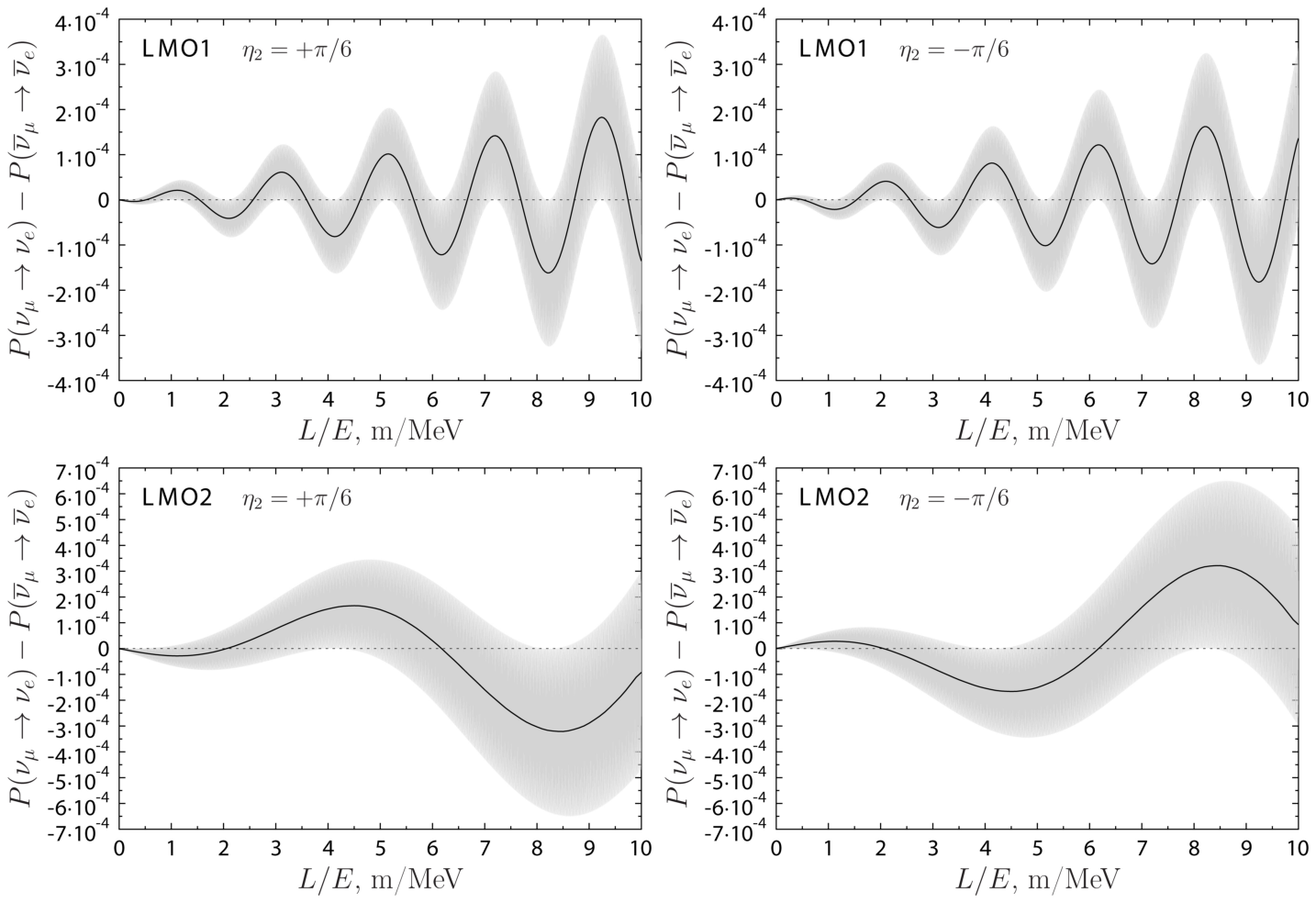}
\caption{The difference between the probabilities of the appearance of
electron neutrinos and antineutrinos for cases of Fig.~\ref{Fig_LMO1} (upper
panels, LMO1) and Fig.~\ref{Fig_LMO2} (lower panels, LMO2) versus the ratio of
the distance $L$ from the source to the neutrino energy $E$ in the beams of
muon neutrinos and antineutrinos, at $\eta_2=+\pi/6$ (left panels) and
$\eta_2=-\pi/6$ (right panels). The gray region corresponds to exact
calculations due to fast oscillations caused by the presence in the model of
fifth sterile neutrino with mass of the order of $1$~{\rm keV}, while the
solid curves show probability values averaged over small-scale spatial
oscillations.}
\label{Fig_Diff}
\end{figure}

In the framework of our model the probabilities of the appearance of electron
neutrinos and antineutrinos in accelerator beams of muon neutrinos and
antineutrinos as a function of the ratio of the distance $L$ to the neutrino
energy $E$ are shown in Figs.~\ref{Fig_LMO1}--\ref{Fig_LMO2}, respectively,
for the mixing matrix in the form considered in this paper
(equations~(\ref{eq_Utilde})--(\ref{eq_matricesab})) and for the two mass
options, namely, LMO1 (equation (\ref{eq_LMO1})) and LMO2
(equation (\ref{eq_LMO2})).

Figure~\ref{Fig_LMO1} shows the appearance probabilities of $\nu_e$ (left
panels) and $\overline{\nu}_e$ (right panels) in the beams of $\nu_{\mu}$ and
$\overline{\nu}_{\mu}$, respectively, as a function of the ratio of the
distance $L$ to the neutrino energy $E$ at the value of the parameter
$\epsilon=0.01$, for the parameter $\eta_2=+\pi/6$ (upper panels) and
$\eta_2=-\pi/6$ (lower panels), and for the LMO1 case (equation~\ref{eq_LMO1})
of neutrino mass distribution. Due to presence in the model of fifth neutrino
with mass of the order of $1$~{\rm keV}, the exactly calculated curves are
fast-oscillating functions of $L/E$ parameter with a smoothly oscillating
envelopes that results to a grey region in Fig.~\ref{Fig_LMO1}. After
averaging over these fast oscillations (solid curves in Fig.~\ref{Fig_LMO1})
that is quite reasonable from point of view of the experiment, the
contribution of sterile neutrinos has the character of smooth oscillations.
Furthermore, at $\eta_2=+\pi/6$ (left upper panel) these oscillations for
neutrinos are in phase with the oscillations, which are obtained with the help
of the standard formula of the (3+1) model for both the probability
$P(\nu_{\mu}\rightarrow\nu_e)$ and
$P(\overline{\nu}_{\mu}\rightarrow\overline{\nu}_e)$, that is by the formula
$P=\sin^2(2\theta)\sin^2(1.27\Delta m^2_{41}L/E)$, where $L$ is the distance
to the detector in m, $E$ is the energy of neutrinos in MeV, and
$\Delta m^2_{41}$ is the neutrino mass-squared difference in ${\rm eV^2}$. On
the other hand, for the same value of $\eta_2=+\pi/6$ it is not the case for
antineutrino oscillations (right upper panel), where the (3+1) oscillations
have a phase shift with respect to the (3+1+2) oscillations. The situation is
inverted at $\eta_2=-\pi/6$ (lower panels). So we have the essential
difference between oscillations in the frameworks of these models due to the
additional source of CP violation in the (3+1+2) model. This property of
oscillations in the model with several neutrinos can be of important practical
consequence while processing experimental data.

In Figure~\ref{Fig_LMO2}, the results for $P(\nu_{\mu}\rightarrow\nu_e)$
(right panels) and $P(\overline{\nu}_{\mu}\rightarrow\overline{\nu}_e)$ (left
panels) are shown as a function of the ratio of the distance $L$ to the
neutrino energy $E$ at the value of the parameter $\epsilon=0.02$, for the
parameter $\eta_2=+\pi/6$ (upper panels) and $\eta_2=-\pi/6$ (lower panels),
and for the LMO2 case (equation~\ref{eq_LMO2}) of neutrino mass distribution.
The correlation between the (3+1) and (3+1+2) oscillations is the same as in
Figure~\ref{Fig_LMO1}. There is the visual difference between the averaged
probabilities (solid curves) of the appearance of neutrinos and antineutrinos.
The scale of the effect for Fig.~\ref{Fig_LMO1} and Fig.~\ref{Fig_LMO2}
differs several times in magnitude. Note that one light sterile neutrino with
mass value of the order of $1$~eV is sufficient to explain the possible
accelerator anomalies in neutrino data within the framework of the model
considered in this paper.

Graphs of the difference of the averaged probabilities of appearance of
electron neutrinos and antineutrinos (asymmetry) for the LMO1 and LMO2 cases
are given in the upper and lower panels of Fig.~\ref{Fig_Diff}, respectively,
for two values of the parameter $\eta_2=\pm\pi/6$ and for the parameter
$\epsilon$ values that correspond to those in Figs.~\ref{Fig_LMO1} and
\ref{Fig_LMO2}. An important result is an oscillating change of sign of this
difference versus the $L/E$ value. Note that the sign of the asymmetry changes
also when the sign of the parameter $\eta_2$ changes. When the neutrino energy
increases (and $L/E$ decreases), detection of the asymmetry in the neutrino
and antineutrino yields becomes much more difficult.

All these results are characteristic features of the considered version of
(3+1+2) model of active neutrinos with allowance made for SN contributions,
and they can be used for interpreting the available experimental data and
predicting the results of new experiments related to the problem of existence
of sterile neutrinos.

\subsection{Estimates of some model parameters on the basis of currently
available experimental data}
\label{Subection_ExperimentalValuesOfModelParameters}

\begin{figure}
\includegraphics[width=0.75\textwidth]{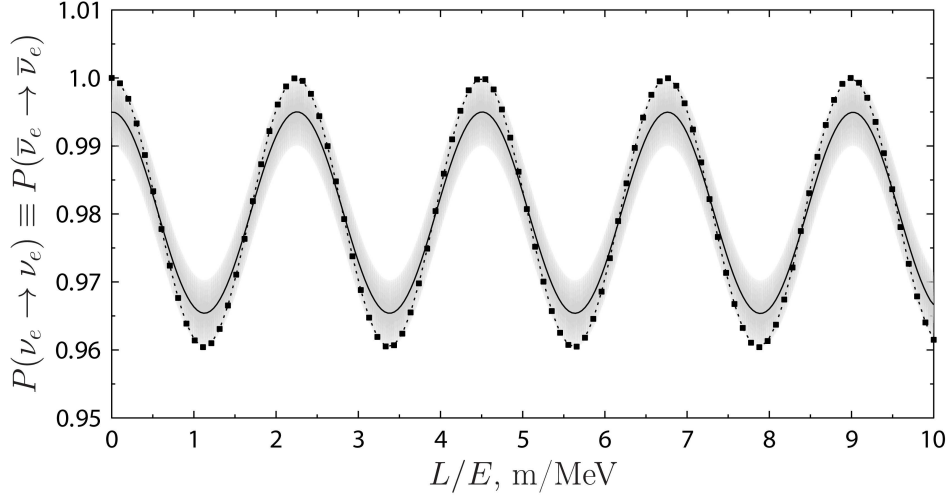}
\centering
\caption{The survival probability of electron neutrinos and antineutrinos
versus the ratio of the distance $L$ from the source to the neutrino energy
$E$ in the beams of electron neutrinos and antineutrinos. The value
$\epsilon=0.005$ of the coupling constant of active and sterile neutrinos is
taken for the case of the mixing matrix considered in this paper
(equations~(\ref{eq_Utilde})--(\ref{eq_matricesab})) with $\eta_2=\pi/6$.
The difference of masses squared with the light SN is equal to
$\Delta m^2_{41}=1.1~{\rm eV}^2$. The gray region arises as a result of exact
calculations due to fast oscillations caused by the presence in the model of
fifth sterile neutrino with mass of the order of $1$~{\rm keV}, while the
solid curve show probability values averaged over small-scale spatial
oscillations. The dashed curve with black squares shows probability values
calculated in the two-level approximation of the (3+1) model and the
two-neutrino mixing with $\sin^2(2\theta_{\rm ee})=0.0396$ and
$\Delta m^2_{41}=1.1~{\rm eV}^2$ from the global fit of the reactor
antineutrino anomaly and gallium neutrino anomaly experimental data
\cite{Dentler}.}
\label{Fig_EXP1}
\end{figure}
\begin{figure}
\includegraphics[width=0.75\textwidth]{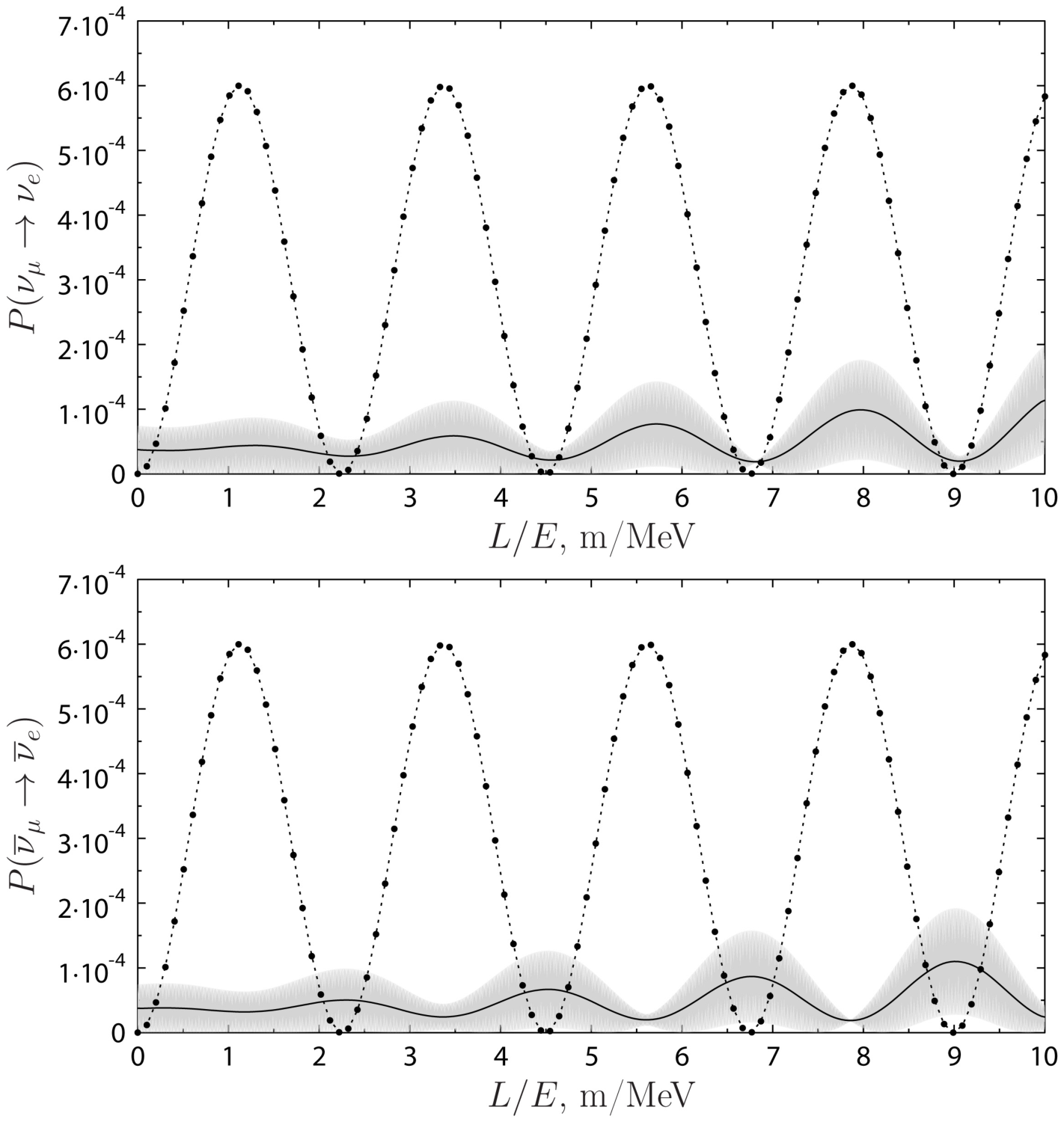}
\centering
\caption{The probability of appearance of electron neutrinos (upper panel) and
antineutrinos (lower panel) versus the ratio of the distance $L$ from the
source to the neutrino energy $E$ in the beams of muon neutrinos and
antineutrinos, respectively. The value $\epsilon=0.005$ of the coupling
constant of active and sterile neutrinos is taken for the case of the mixing
matrix considered in this paper
(equations~(\ref{eq_Utilde})--(\ref{eq_matricesab})) with $\eta_2=\pi/6$. The
difference of masses squared with the light SN is equal to
$\Delta m^2_{41}=1.1~{\rm eV}^2$. The gray region arises as a result of exact
calculations due to fast oscillations caused by the presence in the model of
fifth sterile neutrino with mass of the order of $1$~{\rm keV}, while the
solid curves show probability values averaged over small-scale spatial
oscillations. The dashed curves with black circles show probability values
calculated in the two-level approximation of the (3+1) model and the
two-neutrino mixing with $\sin^2(2\theta_{\rm \mu e})=0.0006$ and
$\Delta m^2_{41}=1.1~{\rm eV}^2$ from the global fit of the accelerator
anomaly experimental data \cite{Dentler}.}
\label{Fig_EXP2}
\end{figure}

At present, experiments are carried out to make sure that SBL neutrino
anomalies exist and can be explained due to the effects of one or a few SN
\cite{Dentler,Giu,Kang}. The simple (3+1) scheme, as a rule, is used for
explanation of SBL anomalies, which includes three AN and one SN with a mass
value of the order of 1~{\rm eV}. In this case, the approximation with
oscillations between only two neutrino mass states (the two-level
approximation) is applied for calculation of probabilities of appearance and
disappearance of AN with $\Delta m^2_{41}$ as a characteristic parameter. In
the framework of this approximation, the probabilities of neutrino or
antineutrino transitions are equal to each other. In our more complicated
model the asymmetry between the probabilities of transitions of muon neutrinos
to electron neutrinos and muon antineutrinos to electron antineutrinos take
place (see Fig.~\ref{Fig_Diff}). And so one of the main conclusions of the present paper is that the processing of experimental data for the neutrino transitions should be done separately from the antineutrino transitions.

There is considerable evidence for the SBL neutrino anomalies and their
interpretation on the basic of the SN hypothesis (see, e.g., \cite{Giu,Kang}).
For a comparison with the available experimental data we prefer to use the
updated global analysis of neutrino oscillations in the presence of eV-scale
SN that has been done in Ref.~\onlinecite{Dentler}. The results for
$\Delta m^2_{41}$, $\sin^2(2\theta_{\rm ee})$ and
$\sin^2(2\theta_{\rm \mu e})$ obtained in this global fit are used below for
estimations of the light SN mass and the parameter $\epsilon$ by comparison
with the results of numerical model calculations. Other model parameters will
take test values as in the previous
Section~\ref{Subection_TestValuesOfModelParameters}.

Figure~\ref{Fig_EXP1} shows the survival probabilities of both $\nu_e$ and
$\overline{\nu}_e$ in the beams of $\nu_{e}$ and $\overline{\nu}_{e}$,
respectively, as a function of the ratio of the distance $L$ to the neutrino
energy $E$ for the data obtained in the experiments for verification of the
reactor and gallium anomalies. The interpolation of the experimental data with
two-level approximation is given in the Fig.~\ref{Fig_EXP1} by the standard
(dashed) curve marked off by the black squares. The parameters
$\Delta m^2_{41}$ and $\sin^2(2\theta_{\rm ee})$ of this curve are picked out
of ``the domain with the star" on Figure~3 from Ref.~\onlinecite{Dentler}. The
other curves in Fig.~\ref{Fig_EXP1} of the present paper are obtained by
calculations in the framework of our model with
$\Delta m^2_{41}=1.1~{\rm eV}^2$ and $\epsilon=0.005$.

Figure~\ref{Fig_EXP2} shows the transition probabilities of $\nu_{\mu}$ in
$\nu_e$ (upper panel) and $\overline{\nu}_{\mu}$ in $\overline{\nu}_e$ (lower
panel) in the beams of $\nu_{\mu}$ and $\overline{\nu}_{\mu}$, respectively,
for the data obtained in the experiments for verification of the accelerator
anomaly. The interpolation of the experimental data with the two-level
approximation is given in Fig.~\ref{Fig_EXP2} with the standard (dashed) curve
marked off by the black circles. The parameters $\Delta m^2_{41}$ and
$\sin^2(2\theta_{\rm \mu e})$ of this curve are picked out of the admissible
region in the vicinity of ``the combined domain" on Figure~4 (right panel)
from Ref.~\onlinecite{Dentler} with the decay-at-rest and decay-at-flight
data. The other curves in Fig.~\ref{Fig_EXP2} of the present paper are
obtained by our model calculations with the same values of parameters
$\Delta m^2_{41}$ and $\epsilon$ as it was for the model curves on
Fig.~\ref{Fig_EXP1}.

When comparing the results presented in Fig.~\ref{Fig_EXP1} with the results
presented in Fig.~\ref{Fig_EXP2}, it is apparent that the experimental data on
the reactor and gallium SBL neutrino anomalies are represented well by the
model calculations with the selected parameters while it is not the case for
the accelerator SBL neutrino anomaly. In the latter case one can see the
discrepancy between the experimental and the model calculated data. We
interpret this point as the flaw in the data processing for the accelerator
anomaly. This can be due in part to the joint processing of the neutrino and
the antineutrino data. Besides, another problem is a significantly small value
of the probability of transition $\nu_{\mu}(\overline{\nu}_{\mu})$ to
$\nu_e(\overline{\nu}_{e})$, as is seen on the  model curves in
Fig.~\ref{Fig_EXP2}.

\section{Discussion and conclusions}
\label{Section_Conclusion}

In this paper, we considered the phenomenological (3+1+2) neutrino model with
three active and three sterile neutrinos and examined the oscillation
characteristics of active neutrinos in vacuum. The properties of these
characteristics at the test values of the model parameters are numerically
investigated. All calculations were performed for the case of a normal
hierarchy of the mass spectrum of active neutrinos with allowance made for the
possible violation of the CP-invariance in the lepton sector and for the value
$-\pi/2$ of the Dirac CP-phase in the $U_{\rm PMNS}$ matrix. Graphical
dependences of the probabilities of appearance of electron neutrinos and
antineutrinos as a function of the ratio of the distance from the source of
muon neutrinos and antineutrinos to the neutrino energy are given within two
versions of the (3+1+2) model, which differ in neutrino masses, namely, LMO1
and LMO2 (see Figs.~\ref{Fig_LMO1} and \ref{Fig_LMO2}). Besides, graphical
representations of both survival and appearance probabilities of electron
neutrinos and antineutrinos are given with the values of some model parameters
obtained on the basis of comparison with the available experimental data
(see Figs.~\ref{Fig_EXP1} and \ref{Fig_EXP2}).

The results obtained make it possible to interpret the experimental data on
oscillations of neutrinos that admit the existence of the reactor and gallium
anomalies (see Fig.~\ref{Fig_EXP1}) and to a lesser extent of the accelerator
anomaly (see Fig.~\ref{Fig_EXP2}). But the great advantage of the considered
model with regard to the accelerator anomaly is revelation of the asymmetry
between the probabilities of the appearance of electron neutrinos and
antineutrinos (see Fig.~\ref{Fig_Diff} and Fig.~\ref{Fig_EXP2}), which arises
from the specific structure of the mixing matrix between AN and SN (see
equations~(\ref{eq_Umix}) and (\ref{eq_Utilde})). It is important that
sufficiently large values of this asymmetry can be obtained with only one
light SN within the LMO1 and LMO2 cases of the considered model. We note that
the oscillatory character of the acceleration anomaly depends on the value of
the lowest mass of SN. For example, if it is of the order of $1$~eV, then
there are oscillations at short distances. Results obtained for the
probabilities of the appearance of electron neutrinos and antineutrinos (see
Fig.~\ref{Fig_EXP2}) show the possible problems associated with observation of
the LSND and MiniBooNE anomalies, in particular, the problem relevant to the
($\nu$ - $\overline{\nu}$) asymmetry and the problem of a substantially small
value of the transition probability of muon (anti)neutrinos in electron
(anti)neutrinos. The selected value of the parameter $\epsilon$ can explain
the reactor and gallium anomalies, as is seen in Fig.~\ref{Fig_EXP1}, but it
is not the case for the acceleration anomaly (see Fig.~\ref{Fig_EXP2}). In a
latter case, it would be a rather higher value of the parameter $\epsilon$.
Notice that the gallium anomaly is free of the uncertainties of the neutrino
energy spectrum.

In the near future, a number of ground-based experiments are planned, which
are aimed at the search for sterile neutrinos
\cite{Abazajian2012,Gorbunov2014,Gav2017,Gav2018}. 
Let us list some recent experimental results on the SN search.
Two recent results with values of SN oscillation parameters,
 which both differ from results obtained in the other measurements and these results also differ between themselves. The NEUTRINO-4 result \cite{Ser2015} is 
$ \sin^2(2\theta_{ee})=0.39, \, \Delta m_{41}^2 = 7.3 eV^2$. The MiniBooNE result \cite{Agu2018} is 
$\sin^2(2\theta_{\mu e})=0.84, \, \Delta m_{41}^2 = 0.039 eV^2$.
However the results of the DANSS and NEOS/Daya Bay experiments  are independent from the theoretical Huber-Mueller flux calculation \cite{Mu2011,Hu2011} and from the 5 MeV bump effect. The best-fit region of the combined analysis \cite{Gari} of the combined fit of the NEOS/Daya Bay  spectral ratio \cite{Ko} and the ratio of the spectra measured at 10.7 and 12.7 meters from the Kalinin reactor in the DANSS  experiment \cite{ale18} is 
$|U_{e4}|^2=0.012\pm 0.003, \, \Delta m_{41}^2 = 1.29 \pm 0.03 eV^2$. These values are close to the values 
of corresponding parameters of the (3+1+2) model used in the present paper.

\appendix

\section{Analytic expressions for
$P(\nu_{\mu}(\overline{\nu}_{\mu})\rightarrow\nu_e(\overline{\nu}_e))$}

Here we give analytic expressions for the transition probabilities
$\nu_{\mu}(\overline{\nu}_{\mu})\rightarrow\nu_e(\overline{\nu}_e)$,
which are obtained with using the formula (\ref{eq_Bilenky}). For convenience,
the complete probability of each such transition is divided into the sum of
partial contributions $P_{ik}$ corresponding to individual contributions with
indexes $i>k$ to the sum of formula (\ref{eq_Bilenky}). One-type contributions
are combined together in the expressions $P_{41+51}$, $P_{42+52}$ and
$P_{43+53}$. In this way,
\begin{equation}
P(\nu_{\mu}(\overline{\nu}_{\mu})\rightarrow\nu_e(\overline{\nu}_e))=
P_{21}+P_{31}+P_{32}+P_{41+51}+P_{42+52}+P_{43+53}+P_{54}\,,
\label{eqA1}
\end{equation}
where
\begin{eqnarray}
P_{21}&=&(1-\epsilon)^4\sin(2\theta_{12})\cos^2\theta_{13}
\left\{\cos(2\theta_{12})\sin\theta_{13}\sin(2\theta_{23})\cos\delta_{CP}
\right.\nonumber\\
&+&\left.\sin(2\theta_{12})\cos^2\theta_{23}-\sin(2\theta_{12})
\sin^2\theta_{13}\sin^2\theta_{23}\right\}\sin^2\Delta_{12}\nonumber\\
&-&a(1-\epsilon)^4\sin(2\theta_{12})\sin\theta_{13}\cos^2\theta_{13}
\sin\theta_{23}\cos\theta_{23}\sin\delta_{CP}\sin(2\Delta_{12})\,,
\label{eqA2}
\end{eqnarray}
\begin{eqnarray}
P_{31}&=&(1-\epsilon)^4\sin(2\theta_{13})\sin\theta_{23}\left\{
\sin(2\theta_{12})\cos\theta_{13}\cos\theta_{23}\cos\delta_{CP}\right.
\nonumber\\
&+&\left.\cos^2\theta_{12}\sin(2\theta_{13})\sin\theta_{23}\right\}
\sin^2\Delta_{13}\nonumber\\
&+&a(1-\epsilon)^4\sin(2\theta_{12})\sin\theta_{13}\cos^2\theta_{13}
\sin\theta_{23}\cos\theta_{23}\sin\delta_{CP}\sin(2\Delta_{13})\,,
\label{eqA3}
\end{eqnarray}
\begin{eqnarray}
P_{32}&=&-(1-\epsilon)^4\sin(2\theta_{13})\sin\theta_{23}\left\{
\sin(2\theta_{12})\cos\theta_{13}\cos\theta_{23}\cos\delta_{CP}\right.
\nonumber\\
&-&\left.\sin^2\theta_{12}\sin(2\theta_{13})\sin\theta_{23}\right\}
\sin^2\Delta_{23}\nonumber\\
&-&a(1-\epsilon)^4\sin(2\theta_{12})\sin\theta_{13}\cos^2\theta_{13}
\sin\theta_{23}\cos\theta_{23}\sin\delta_{CP}\sin(2\Delta_{23})\,,
\label{eqA4}
\end{eqnarray}
\begin{eqnarray}
P_{41+51}&=&-(1-\epsilon)^2(2\epsilon-\epsilon^2)\sin(2\eta_2)\left\{
\cos^2\theta_{12}\sin(2\theta_{13})\sin\theta_{23}\cos\delta_{CP}\right.
\nonumber\\
&+&\left.\sin(2\theta_{12})\cos\theta_{13}\cos\theta_{23}\right\}
(\sin^2\Delta_{14}-\sin^2\Delta_{15})
\nonumber\\
&+&a(1-\epsilon)^2(2\epsilon-\epsilon^2)\sin(2\eta_2)\cos^2\theta_{12}
\sin\theta_{13}\cos\theta_{13}\sin\theta_{23}\sin\delta_{CP}\nonumber\\
&\times&\{\sin(2\Delta_{14})-\sin(2\Delta_{15})\}\,,
\label{eqA5}
\end{eqnarray}
\begin{eqnarray}
P_{42+52}&=&(1-\epsilon)^2(2\epsilon-\epsilon^2)\sin(2\eta_2)\left\{
-\sin^2\theta_{12}\sin(2\theta_{13})\sin\theta_{23}\cos\delta_{CP}\right.
\nonumber\\
&+&\left.\sin(2\theta_{12})\cos\theta_{13}\cos\theta_{23}\right\}
(\sin^2\Delta_{24}-\sin^2\Delta_{25})
\nonumber\\
&+&a(1-\epsilon)^2(2\epsilon-\epsilon^2)\sin(2\eta_2)\sin^2\theta_{12}
\sin\theta_{13}\cos\theta_{13}\sin\theta_{23}\sin\delta_{CP}\nonumber\\
&\times&\{\sin(2\Delta_{24})-\sin(2\Delta_{25})\}\,,
\label{eqA6}
\end{eqnarray}
\begin{eqnarray}
P_{43+53}&=&(1-\epsilon)^2(2\epsilon-\epsilon^2)\sin(2\eta_2)
\sin(2\theta_{13})\sin\theta_{23}\cos\delta_{CP}(\sin^2\Delta_{34}
-\sin^2\Delta_{35})\nonumber\\
&-&a(1-\epsilon)^2(2\epsilon-\epsilon^2)\sin(2\eta_2)\sin\theta_{13}
\cos\theta_{13}\sin\theta_{23}\sin\delta_{CP}\nonumber\\
&\times&\{\sin(2\Delta_{34})-\sin(2\Delta_{35})\}\,,
\label{eqA7}
\end{eqnarray}
\begin{eqnarray}
P_{54}&=&(2\epsilon-\epsilon^2)^2\sin^2(2\eta_2)\sin^2\Delta_{45}\,.
\label{eqA8}
\end{eqnarray}
Here $a=+1$ for neutrinos and $a=-1$ for antineutrinos.

\end{document}